\documentclass[prb,aps,showpacs,twocolumn,amsmath,amssymb]{revtex4}
\usepackage{graphicx}% Include figure files
\usepackage{amsmath}
\begin{document}
\title{Unification of the phonon mode behavior in semiconductor alloys:
\newline
Theory and \textit{ab initio} calculations}
\author{O. Pag\`{e}s}
\email[Author to whom correspondence should be addressed.
Electronic mail:~]{pages@univ-metz.fr}
\author{A.~V. Postnikov}
\author{M. Kassem}
\author{A. Chafi} 
\author{A. Nassour}
\author{S. Doyen}
\affiliation{Laboratoire de Physique des Milieux Denses, 
Universit\'{e} Paul Verlaine, 1 Bd. Arago, 57078 Metz, France}
\date{February 5, 2008}

\begin{abstract}
We demonstrate how to overcome serious problems in understanding 
and classification of vibration spectra in semiconductor alloys, 
following from traditional use of the virtual crystal approximation 
(VCA). We show that such different systems as InGaAs 
(1-bond$\rightarrow$1-mode behavior), InGaP (modified 2-mode) 
and ZnTeSe (2-bond$\rightarrow$1-mode) 
obey in fact the same phonon mode behavior -- hence probably a 
universal one -- of a percolation-type (1-bond$\rightarrow$2-mode). 
The change of paradigm from the `VCA insight' (an averaged microscopic one) 
to the `percolation insight' (a mesoscopic one) offers a promising 
link towards the understanding of alloy disorder. The discussion 
is supported by \textit{ab initio} simulation of the phonon density 
of states at the zone-center of representative supercells at 
intermediary composition (ZnTeSe) and at the impurity-dilute 
limits (all systems). In particular, we propose a simple \textit{ab initio} 
`protocol' to estimate the basic input parameters of our 
semi-empirical `percolation' model for the calculation of the 
1-bond$\rightarrow$2-mode vibration spectra of zincblende alloys. 
With this, the model turns self-sufficient. 
\end{abstract}
\maketitle

\section{Introduction}
When dealing with any kind of mixture the key issue is how to 
handle the substitutional disorder. This generic term covers 
topological disorder, mass disorder, disorder in the bond length, 
disorder in the bond force constant, etc. One can choose either 
a percolation-based approach, which is essentially a mesoscopic 
one, or a microscopically-averaged description. In the first 
case the central notion is the volume fraction of each constituent. 
Some criticality in the dependence of the physical properties 
on the composition of the mixture is then expected at the percolation 
thresholds, where the minor constituent coalesces into a continuum. 
Percolation approaches are typically used for molecular/natural 
mixtures, which are of a forbidding complexity at the microscopic 
scale (statistical arrangement of molecules or grains with a 
distribution of sizes and shapes), and follow thereby the impetus 
given in particular by De Gennes.\cite{01} This approach might seem 
sometimes over-complicated. When facing such kind of `ideally 
disordered' mixtures as conventional semiconductor (SC) alloys 
AB$_{1-x}$C$_x$ [simple atoms distributed at random on a quasi-regular 
lattice, that makes two interpenetrating A and (B,C) fcc sublattices], 
one is tempted to proceed as possibly far with microscopic averaging, 
corresponding then to the virtual crystal approximation (VCA). 
With this, each atom A is surrounded by four virtual B$_{1-x}$C$_x$ 
atoms, i.e. the crystal is viewed at the \textit{macroscopic} scale 
as a continuum. The physical properties are accordingly averaged. 
No singularity is expected in their dependence on $x$. Actually 
the VCA provides, at first sight, a fairly good description of 
many properties in SC alloys. 

In this work we focus on the bond force constant ($K$) in SC alloys. 
$K$ is sensitive to the overall bond distortion (bond length plus 
bond angles) required to accommodate the mismatch in the A--B 
and A--C bond lengths -- a simple rule states `\textit{the closer 
the atoms, the larger $K$}', and as such $K$ provides a sort of integrated 
insight into the substitutional disorder. $K$ is routinely probed 
by Raman and Infrared (IR) spectroscopies via the frequency of 
the transverse optical (TO) phonon. We recall that a transverse (longitudinal) 
optical TO (LO) mode in the Raman/IR spectra corresponds to vibration of the 
rigid A sublattice against the rigid (B,C) sublattice perpendicular 
to (along) the direction of propagation. In a polar crystal a LO mode differs 
from a TO mode in that it carries a coulombian field $\vec{E}$
due to the ionicity of the bond. In a pure crystal this is just 
responsible for an additional restoring force, with the result 
that a LO mode occurs at a higher frequency than a TO mode. In a multi-wave 
system, such as an alloy, the coulombian field $\vec{E}$ is well-known 
to act as a carrier of coherence,\cite{02} that makes LO modes 
a much complicated 
issue. In fact, the individual LO modes with close frequencies do 
$\vec{E}$-couple, which results in a dramatic distortion of the original 
LO lineshapes.\cite{03} In contrast, the individual TO modes do remain 
unaltered,\cite{03} and thereby provide reliable insight into the nature 
(via the TO-frequency) and the population (via the TO-intensity) 
of the individual oscillators present in the crystal. So, in the present work 
we depart from a confusing habit to treat the TO and LO modes on an equal 
footing, and focus most of our attention on TO modes. Below, we emphasize 
accordingly the TO aspect in the different models and approaches used 
throughout the manuscript, to set up a consistent basis for the discussion. 

\begin{figure}[tb]
\centerline{\includegraphics[width=0.48\textwidth]{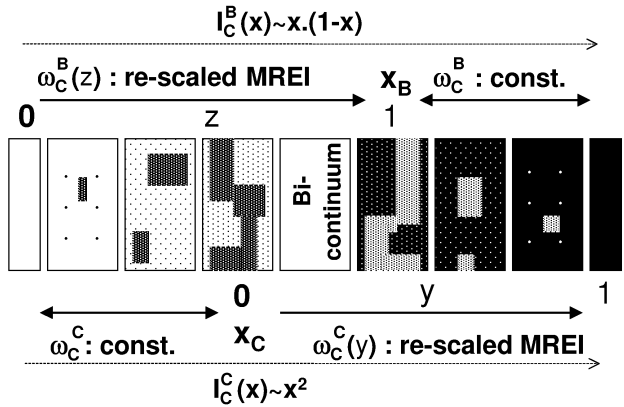}}
\medskip
\caption{%FIG. 1: 
Schematic views of a AB$_{1-x}$C$_{x}$ alloy 
according to the VCA (a) and percolation (b) approaches. 
A simple greyscale code is used. In the alloy the greyscale reinforces 
when the local composition becomes more like that of the corresponding 
pure crystal. A MREI-like correlation with the dependence of the A--C 
TO-frequency ($\omega$) on the actual alloy composition $x$ (a) 
or on re-scaled alloy compositions $y$ and $z$ (b) is emphasized. 
In case (b) critical behaviors occur at the A--C ($x_{\rm C}$=0.19) and 
A--B ($x_{\rm B}$=0.81) bond percolation thresholds. In particular out 
of the percolation regime a minor region forms a dispersion of 
clusters with quasi-stable internal structure, as evidenced by 
a stable grey in scheme (b). This corresponds to a stable phonon 
frequency. The intensity ($I$) aspect is also indicated. For a 
given A--C mode this scales as the total fraction of A--C bond 
in the alloy ($x$) weighted by the scattering volume of the corresponding 
C-rich ($x$) or B-rich ($1-x$) host region. Subscript and superscript 
B (C) refer to the A--B (A--C) bond species and to the pale B-rich 
(dark C-rich) region, respectively. A similar frequency/intensity 
description applies to A--B.}
\end{figure}

The VCA for a bond-related property in a AB$_{1-x}$C$_x$ alloy, such 
as the bond force constant $K$, comes to a picture where the bond 
is immersed into a continuum whose physical properties are smoothly 
dependent on the alloy composition $x$, as schematically represented 
in Fig.~1a. With this, the bonds of like species are all equivalent 
in the alloy, thereby contributing to a unique TO
mode in the Raman/IR spectra. The intensity scales as the corresponding 
fraction of bonds, and when x changes the mode shifts regularly 
between the natural frequency in the pure crystal and the impurity 
frequency. No singularity is expected. Such 1-bond$\rightarrow$1-mode TO 
behavior is well-accounted for by the modified-random-element-isodisplacement 
(MREI) model as worked out by Chang and Mitra in the sixties, 
based on the VCA.\cite{04} In order to grasp the whole behavior, one 
needs only the frequency of the impurity mode, referred to as
$\omega_{\rm imp}$ below, the TO frequency in the pure crystal being normally 
well-known. Now, a short (long) impurity bond is tensed (compressed) 
in a matrix with a large (small) lattice constant, which reduces (enlarges) 
$K$, with concomitant impact on $\omega_{\rm imp}$ (see the `rule' in 
italics above). Still, generally $\omega_{\rm imp}$ remains close 
to the parent TO frequency, because each bond tends to retain its natural bond 
length in an alloy, as is well-known.\cite{05} So, the trend in 
a `TO-frequency vs. $x$' plot should be that the A-B and A-C TO branches 
do remain quasi-parallel, with slight but finite slopes. 

Two main types of TO mode behavior proceed from the MREI-VCA, supporting 
a corresponding classification of the Raman/IR TO data: 
($i$) If the parent TO frequencies are much distinct, then the alloy exhibits 
two well-separated A--B and A--C TO branches, corresponding to a pure 
1-bond$\rightarrow$1-mode (2-mode) behavior. 
($ii$) If the parent TO frequencies are close enough, the A--B and A--C 
TO branches merge into an apparent 2-bond$\rightarrow$1-mode (mixed-mode) 
behavior with a unique (A--B, A--C)-mixed TO mode that has quasi-stable 
intensity, and shifts regularly between the parent TO modes. There seems 
to exist a sort of intermediate-type behavior, the so-called ($iii$) 
modified-two-mode behavior,\cite{06} with two (A--B, A--C)-mixed TO modes, i.e. 
a dominant one of type ($ii$), plus a minor one joining the impurity modes 
that does not allow to fully discard the type ($i$). Note that while
the MREI-VCA model fully accounts for types ($i$) and ($ii$), it fails 
to account for the type ($iii$), even qualitatively. 
Representative systems are In$_{1-x}$Ga$_x$As, ZnTe$_{1-x}$Se$_x$ and 
In$_{1-x}$Ga$_x$P, respectively.\cite{07} The related MREI-VCA 
TO schemes are schematically represented in Fig.~2, while approximating 
the MREI branches to straight lines, for simplicity. The schemes 
were built up by using the traditional sets of impurity frequencies: 
InAs:Ga$\sim$241~cm$^{-1}$ and GaAs:In$\sim$237~cm$^{-1}$ 
(Ref.\onlinecite{08}), 
InP:Ga$\sim$347~cm$^{-1}$ and GaP:In$\sim$390~cm$^{-1}$ 
(Ref.\onlinecite{09}), and 
ZnTe:Se$\sim$159~cm$^{-1}$ and ZnSe:Te$\sim$116~cm$^{-1}$ 
(Ref.\onlinecite{10}), 
with the usual notation. In ZnTeSe it is commonly admitted 
that the impurity modes are screened by the disorder-induced 
phonon density of states, and thereby do not give rise to any 
observable feature in the Raman/IR spectra.\cite{10}
Note that in the MREI--VCA schemes related to InGaP and ZnTeSe,
the basic TO branches, as obtained simply by joining the corresponding
parent and impurity modes, cross each other (not shown), which contradicts
the expected trend of quasi-parallel TO branches.

\begin{figure*}[tb]
\centerline{\includegraphics[width=0.95\textwidth]{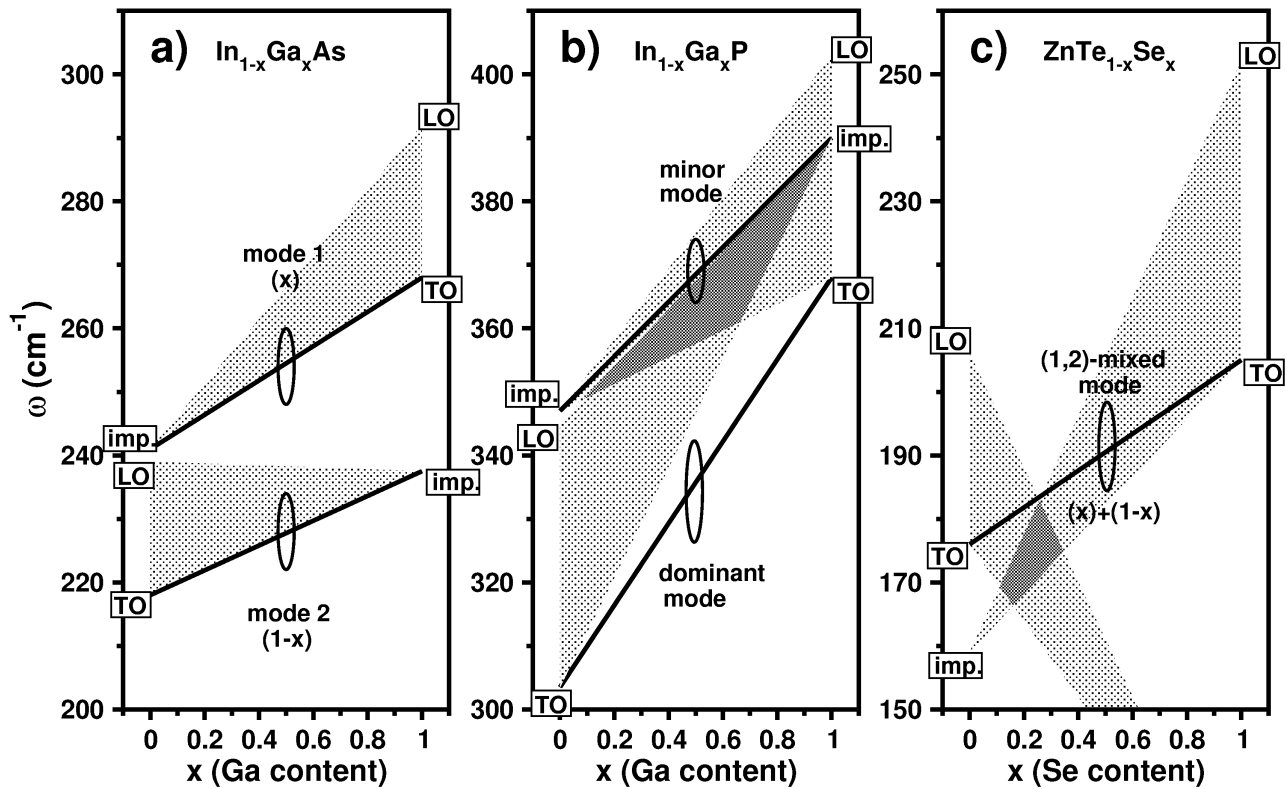}}
\medskip
\caption{%FIG. 2:
Simplified 1-bond$\rightarrow$1-mode TO (thick lines) 
MREI-VCA schemes of InGaAs (a), InGaP (b) and ZnTeSe (c). Only those 
TO modes actually observed in the Raman/IR spectra are represented. 
The intensity of each TO mode basically scales like the corresponding
bond fraction (refer to Fig.~1a), as specified within brackets (when
this is known). The optical bands used for the Elliott-CPA criterion 
are shown as shaded areas. The $\omega_{\rm imp}$ values for InGaAs, 
InGaP and ZnTeSe are taken from Refs.~\onlinecite{08}, 
\onlinecite{09} and \onlinecite{10}, respectively.}
\end{figure*}

Elliott \textit{et al.}\cite{11} worked out a theoretical criterion within 
the coherent potential approximation (CPA) to distinguish between 
type ($i$) and type ($ii$). As the CPA is well-suited mainly for 
the treatment of low concentration of simple defects in otherwise 
perfect media, the Elliott-CPA criterion aims at deriving the 
whole phonon mode behavior of an alloy from its behavior at the 
dilute limits. The criterion predicts type ($ii$) when the A--B and A--C 
TO--LO bands do overlap in the alloy, type ($i$) other\-wise. Here the 
optical bands are simply built up by linear convergence of the 
parent TO and LO frequencies in the pure crystal onto the related 
impurity frequency. 

Generally the MREI-VCA (experimental) and Elliott-CPA (theoretical) 
classifications were found to be remarkably consistent, and an 
exhaustive review of the phonon mode behavior of many ternary 
alloys in terms of types ($i$) and ($ii$) was proposed by Taylor.\cite{12} 
A remarkable exception is InGaP, whose special Raman/IR behavior 
is not covered by the Elliott-CPA criterion. In fact, the Elliott-CPA 
criterion predicts a type ($i$) behavior for InGaP, in apparent 
contradiction with the experimental findings that seem to indicate 
a strong overlapping of the optical bands in GaInP (see the over-shaded
area in Fig.~2b). However, 
careful analysis of the Raman/IR data reveals that even the representative 
alloys we discuss do not fit into the MREI-VCA/Elliott-CPA classification. 
Indeed, there are three features in the TO Raman spectra of InGaAs, rather 
than two (refer to Fig.~3 in Ref.~\onlinecite{08}). Also, the IR spectra 
of ZnTeSe exhibit two clear resonances, not only one 
(refer to Fig.~1 in Ref.~\onlinecite{13}). At last, 
the minor TO mode of InGaP was identified by \textit{ab initio} calculations 
as a pure Ga--P mode, not as a (In--P, Ga--P)-mixed mode (refer to 
Fig.~3a in Ref.~\onlinecite{14}).

Our view is that the VCA misses the essence of the phonon behavior 
in alloys. Recognizing the local character of $K$ we conclude that 
the proper understanding of TO phonons requires detailed insight into 
the topologies of the (B,C)-substituting species, which falls into the scope 
of the percolation site theory.\cite{15} Precisely, the apparent anomalies 
in InGa(As,P) could be explained within a 1-bond$\rightarrow$2-mode 
percolation model,\cite{03} introducing a description of a random alloy 
at the \textit{mesoscopic} scale, in terms of a composite made of the B-rich 
and C-rich regions, both resulting from natural fluctuations in the alloy 
composition at the local scale. The whole picture is summarized in Fig.~1b. 
For a given bond, each region 
brings a specific TO mode, the result of different local bond distortions. 
In a `TO-frequency vs. $x$' plot, this leads to a splitting of each original 
MREI-like TO branch into a symmetrical double-branch attached at its two 
ends to the parent and impurity modes. The intensity of each 
`sub-mode' scales as the total fractions of like bonds in the 
alloy, i.e. as $(1-x)$ for A--B and $x$ for A--C, weighted by the scattering 
volume of the corresponding B-rich ($1-x$) or C-rich ($x$) host region. 
In particular, singularities in the TO -frequency occur at the bond 
percolation thresholds, where the minor A--B ($x_{\rm B}$=0.19) and A--C 
($x_{\rm C}$=0.81) bonds become connected throughout the crystal. This is 
absent with the MREI-VCA. Below the percolation threshold, the minor region 
consists of a dispersion of finite clusters with similar internal structures, 
which generates quasi-invariance in the TO -frequency (fractal-like regime). 
Above the percolation threshold, the finite clusters coalesce into a continuum 
whose local composition turns smoothly $x$-dependent, as does the 
TO frequency (normal regime). There, a re-scaled MREI approach 
applies where the continuum is viewed as a pseudo-ternary alloy, 
and takes an apparent composition that varies from 0 to 1 over 
its domain of existence (refer to ``y'' and ``z'' in Fig.~1b). Basically, 
two adjustable parameters are required per bond to figure out 
the whole behavior, i.e. $\omega_{\rm imp}$ -- as for the MREI model, 
plus the splitting $\Delta$ between like TO modes on the onset of the 2-mode 
behavior just departing from the impurity limit. 

In this work we investigate whether, among the usual SC alloys, 
the percolation scheme for InGa(As,P) is fortuitous, i.e. contingent 
upon the (Ga,In)-substituting species for some reason, or reflects 
a deeper -- universal -- reality by extending further to the representative 
system of the remaining class ($ii$), i.e. ZnTeSe. The perspective 
is a change of paradigm regarding the way to describe phonons 
in alloys in general. 

The manuscript is organized as follows. In Sec.~II we re-examine 
representative sets of optical phonon frequencies related to 
ZnTe$_{1-x}$Se$_x$ taken in the literature. We conclude to a three-oscillator 
type behavior in the Raman/IR spectra, that fits into 
the 1-bond$\rightarrow$2-mode percolation scheme. A suitable version 
of the percolation model is derived, supported by independent insight 
into the impurity modes via existing extended X-ray Absorption Fine Structure 
(EXAFS) measurements of the impurity bond lengths. The (TO, LO) Raman 
lineshapes are derived in Sec.~III for a clear overview 
of the phonon mode behavior of ZnTeSe. \textit{Ab initio} insight 
into the zone-center TO phonon density of states (ZC TO-DOS), that mimics 
the Raman signal, is produced at the stoichiometry ($x$=0.5) for comparison.
In Sec.~IV, we present a unified description of the phonon mode behavior 
in SC alloys within the percolation scheme, and confront it with the previous 
MREI-VCA/Elliott-CPA classification. At last, in Sec.~V, we propose 
a simple \textit{ab initio} `protocol' to estimate the input parameters 
($\omega_{\rm imp.}$, $\Delta$) of the semi-empirical percolation model. 

\section{PERCOLATION PICTURE FOR 
Z\lowercase{n}T\lowercase{e}$_{1-x}$S\lowercase{e}$_x$}
A priori the prerequisites for the detection of a possible 
1-bond$\rightarrow$2-phonon behavior seem favorable in ZnTe$_{1-x}$Se$_x$. 
Indeed the lattice mismatch in ZnTeSe is similar to that of InGa(As,P), i.e. 
$\sim$7\%, so that similar lattice distortions can be expected, and thereby 
also similar phonon properties. Further, the dispersion of the 
TO mode is small both in ZnSe ($\sim$6~cm$^{-1}$) and 
ZnTe ($\sim$8~cm$^{-1}$),\cite{16} so that even small differences in the local 
bond distortions may generate fluctuations in the TO frequency greater than 
the dispersion, thereby giving rise to phonon localization, i.e. 
to well-separated TO modes in the Raman/IR spectra.\cite{17} In fact, already 
three decades ago Artamonov \textit{et al.}\cite{18} mentioned 
an `evident analogy' between the phonon behaviors of InGaP and ZnTeSe. 
Later on, Yang \textit{et al.}\cite{19} proposed to reclassify ZnTeSe as 
a type ($iii$) system, not as a type ($ii$). Altogether this suggests that
a InGaP-like version of the percolation model should apply to ZnTeSe. 

\begin{figure*}[ht]
\centerline{\includegraphics[width=0.95\textwidth]{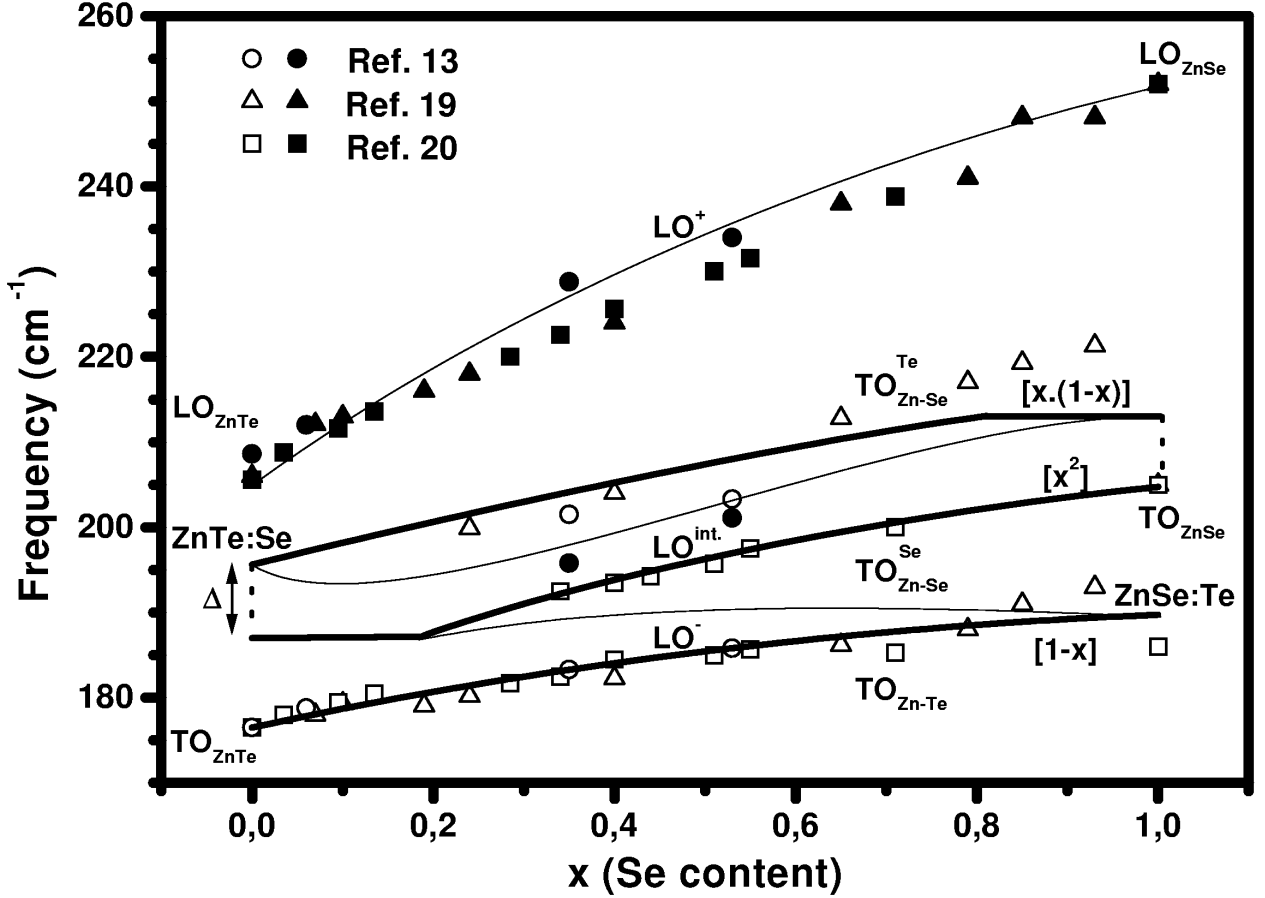}}
\medskip
\caption{%FIG. 3: 
TO (thick lines) / LO (thin lines) percolation picture 
for ZnTe$_{1-x}$Se$_x$. This is built up from representative sets of 
TO (open symbols) / LO (filled symbols) frequencies taken from the literature,
as indicated. The intensity of each TO mode scales as the corresponding bond 
fraction (see Fig.~1b), as specified within square brackets. 
ZnTe:Se and ZnSe:Te refer to 
the impurity modes of Se and Te in ZnTe and ZnSe, respectively. 
Dotted lines indicate the onset of the ZnSe-like 1-bond$\rightarrow$2-mode 
TO behavior just departing from the dilute limits.}
\end{figure*}

Representative sets of (TO, LO) frequencies in ZnTeSe obtained from 
Raman/IR analysis, as taken in the literature,\cite{13,19,20} are displayed 
in Fig.~3. As usual\cite{03} in the discussion of the phonon mode behavior, 
we focus on the TO modes, whereas the discussion on LO modes follows, among 
some additional details, in Sec.~III. The data indicate three equally spaced 
TO modes in the range 175--215~cm$^{-1}$. We propose a three-oscillator 
[1$\times$(Zn--Te), 2$\times$(Zn--Se)] version of the percolation model, 
with a unique TO$_{\rm Zn-Te}$ branch below a well-resolved Zn-Se 
TO double-branch. Detail is given further on. In particular 
the percolation scheme 
incorporates a general trend that bonds are longer (shorter) 
in environments rich of the shorter (longer) bond\cite{04} -- as also 
evidenced by Silverman \textit{et al.}\cite{21} in InGaP by using 
first-principles calculations -- which reduces (enlarges) the TO frequency. 
So, the middle and upper TO branches in Fig.~3 are attributed to 
Zn--Se vibrations in the Se- and Te-rich regions, and are 
accordingly labeled as TO$_{\rm Zn-Se}^{\rm Se}$ and TO$_{\rm Zn-Se}^{\rm Te}$. 
The ($\omega_{\rm imp}$, $\Delta$) values for the Zn--Te and Zn--Se bonds are 
estimated as ($\sim$189~cm$^{-1}$, $\sim$0~cm$^{-1}$) and 
($\sim$195~cm$^{-1}$, $\sim$8~cm$^{-1}$), respectively, as explained
hereafter.

The $\omega_{\rm imp}$ values were derived by relying on thorough 
EXAFS measurements of bond length in ZnTeSe.\cite{22} The shift 
$\Delta \omega_{\rm T}^2$ in the square TO frequency of an impurity mode 
with respect to the corresponding TO mode in the pure crystal was inferred 
from the related difference in bond length $\Delta l$ via the TO mode 
Gr\"{u}neisen parameter $\gamma_{\rm T}$ of the pure crystal, by using 
the simple relation\cite{08}
\begin{equation}
\frac{\Delta \omega_{\rm T}^2}{\omega_{\rm T}^2} =
- 6\gamma_{\rm T}\frac{\Delta l}{l}\,.
\end{equation}
Denominators refer to the pure crystal. For the $\Delta l$
and $\gamma_{\rm T}$ values related to the Zn--Se (Zn--Te) bond, we have
taken 2.600--2.643~{\AA} (2.480--2.452~{\AA}) and 1.7 (1.4) according to 
Refs.~\onlinecite{22} and \onlinecite{23}, respectively. 
For each bond species, the impurity bond length was determined by
linear adjustment of the whole `bond length vs. $x$' EXAFS dependence,
and extrapolation to the dilute limit. The whole procedure
was earlier tested with InGaP, with much success.\cite{03}

Now we turn to $\Delta$. The lower TO-data set is attached to the limit 
($x{\sim}$0,1) Zn--Te frequencies at its two ends, indicating an apparent 
1-bond$\rightarrow$1-mode MREI behavior for the Zn--Te bond, which comes to
$\Delta_{\rm Zn-Te}{\sim}$0~cm$^{-1}$. The two remaining TO-data sets share 
quasi-symmetrically on each side of the limit Zn--Se frequencies, indicating
a well-resolved 1-bond$\rightarrow$2-mode type behavior for the Zn-Se bond, 
i.e. a finite $\Delta_{\rm Zn-Se}$ value. This was adjusted to 
$\sim$8~cm$^{-1}$ by fitting a rescaled-MREI curve to the data in the continuum
regime of the dominant TO$_{\rm Zn-Se}^{\rm Se}$ branch at large Se content. 
In this composition range the TO$_{\rm Zn-Se}^{\rm Se}$ mode shows up
strongly in the Raman/IR spectra (see next Sec.), so that its frequency
could be determined with high accuracy.
The remaining pieces of the Zn--Se double-branch, i.e. the two dispersion 
regimes (fractal-like) plus the rescaled-MREI curve in the continuum regime 
(normal) of the TO$_{\rm Zn-Se}^{\rm Te}$ branch, were directly inferred 
by symmetry. 

In Fig.~3 both the TO (thick lines) and LO (thin lines) theoretical curves 
are in good agreement with the data, in spite of the simplicity of our model. 
Slight discrepancy exists with respect to the TO$_{\rm Zn-Se}^{\rm Te}$
branch in the Te-dilute limit (less than 8~cm$^{-1}$), where the 
mode becomes weak and hard to detect (see next Sec.). 

\section{(TO, LO) Raman lineshapes}
\begin{figure*}[t] 
\centerline{\includegraphics[width=0.9\textwidth]{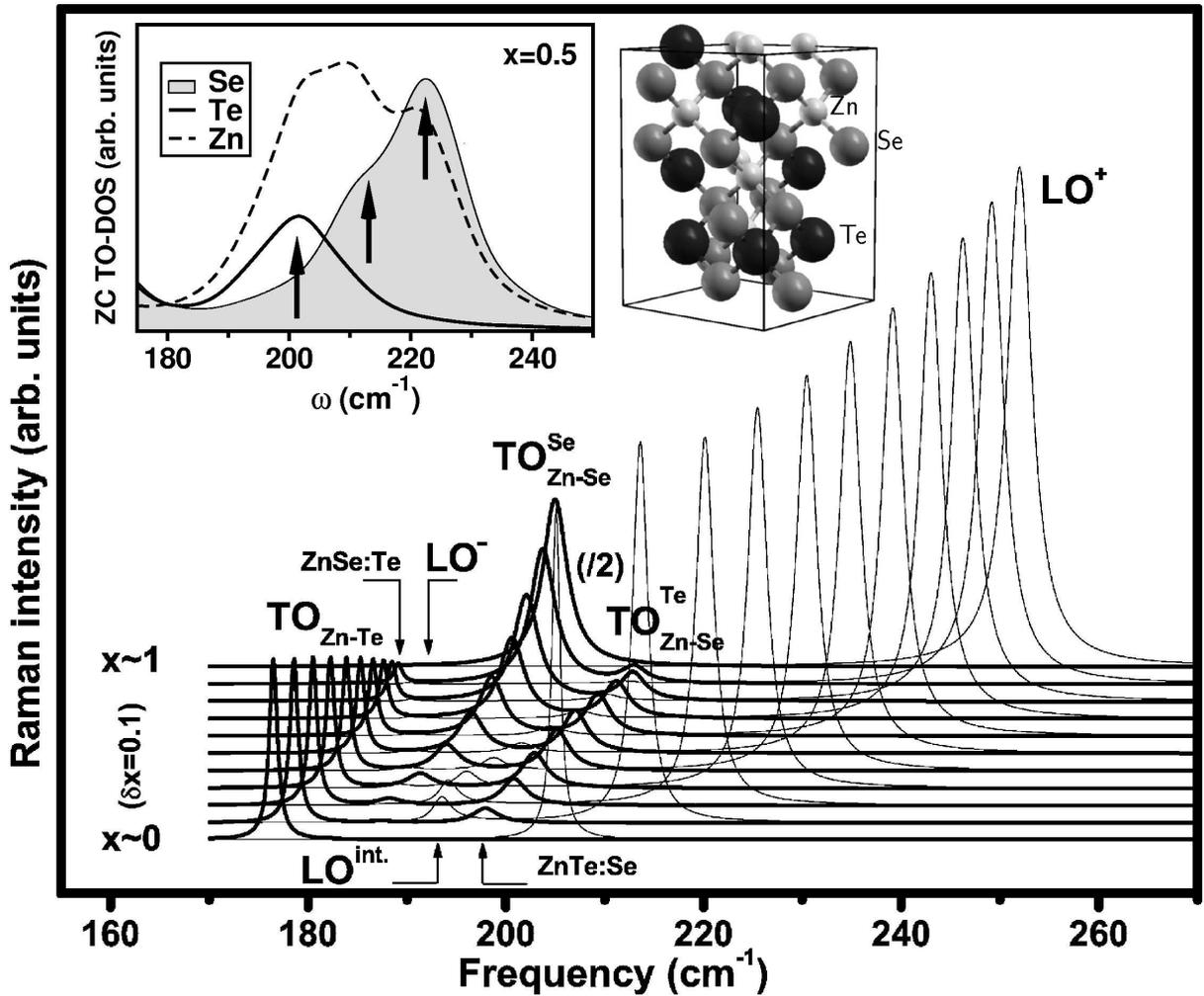}}
\medskip
\caption{%FIG. 4:
TO (thick lines) and LO (thin lines) Raman lineshapes 
of ZnTe$_{1-x}$Se$_x$. These were calculated from the frequency/intensity
TO information displayed in Fig.~3 by using Zn--Se (3~cm$^{-1}$) 
and Zn--Te (1~cm$^{-1}$) phonon dampings that scale like in the 
pure crystals. The bottom LO mode is divided by 2, as indicated. The 
ZC TO-DOS per atom for uniformly-damped Zn, Se and Te oscillators 
(10~cm$^{-1}$) at $x{\sim}$0.5 is shown in the inset for comparison, 
together with the corresponding quasi-random supercell. There, 
the arrows indicate a three-oscillator behavior.}
\end{figure*}

In Fig.~3 all the ingredients are there to calculate the TO and LO
Raman lineshapes of random ZnTe$_{1-x}$Se$_x$. We use the generic 
equation,\cite{03}
\begin{equation}
I \propto \mbox{Im} \left\{ -\epsilon_{r}^{-1} 
\left[ 1+\sum_p C_p K_p L'_p \right]^2
+\sum_p C_p^{2} \frac{K_p^2 L'_p }{4\pi Z_p^2 } 
\right\}\,.
\end{equation}
In a form restricted to the second member, Eq.~(2) gives access to the TO
modes, while in full it provides the LO modes. The summation runs over 
the complete collection of oscillators present in the crystal; 
$\epsilon_r$ is the relative dielectric function of the whole crystal. We 
use a classical form generalized to a discrete series of harmonic 
oscillators, i.e. three in the present case. 
$C_p$, $K_p$ and $L'_p$ are: the Faust--Henry coefficient (which measures 
the relative Raman efficiencies of the TO and LO modes), the TO-frequency 
squared, and the classical Lorentzian response of oscillator $p$, respectively. 
$Z_p$ is expressed according to the standard MREI terminology,\cite{04} and 
relates to the oscillator strength $S_p$. Both $C_p$ and $S_p$ scale as 
the fraction of oscillator $p$ in the crystal. They are accordingly derived 
from the ($C_p$, $\epsilon_{\infty}$, TO--LO) values in the pure ZnSe and ZnTe
crystals, taken as ($-$0.56, 5.75, 206--252~cm$^{-1}$) and 
($-$0.32, 7.20, 176--206 cm$^{-1}$), respectively. For sake of consistency 
in the calculations, the Faust--Henry coefficients of pure ZnSe and ZnTe were 
taken from the same source, i.e. corresponding to the ratio of the ionic to 
electronic parts of the static electro-optic effect,\cite{24} as calculated 
by Shih \textit{et al}.\cite{25} A damping term was introduced in 
$L'_p$ so as to reproduce the finite linewidths of the experimental 
Raman/IR lines. Small dampings were used for the Zn--Se (3~cm$^{-1}$) 
and Zn--Te (1~cm$^{-1}$) modes -- for a clear overview of the whole 
collection of individual oscillators. Care was taken that the 
latter dampings scale like in the pure crystals, i.e. in the ratio 3:1, as 
inferred from the full widths at half maximum of the Raman lines 
obtained in similar conditions with pure ZnSe and ZnTe crystals of the same 
generation (see Figs 1 in Refs~\onlinecite{26} and \onlinecite{27}).
The resulting 
three-mode (TO, LO) Raman lineshapes of ZnTe$_{1-x}$Se$_x$ depending on 
the alloy composition $x$ are displayed in Fig.~4. 

First we discuss the LO modes (thin lines in Figs 3 and 4). In ZnTeSe 
the individual LO$_{\rm Zn-Te}$, LO$_{\rm Zn-Se}^{\rm Se}$
and LO$_{\rm Zn-Se}^{\rm Te}$ modes (not shown) do $\vec{E}$-couple strongly 
because the Zn--Se and Zn--Te oscillators are so close, which renders the 
LO situation especially complicated. Basically the available oscillator 
strength, independently of its origin, i.e. ZnSe- or ZnTe-like, 
is almost fully channeled into a `giant' LO$^{+}$ mode at high frequency 
that has quasi-stable intensity and shifts regularly between the 
LO modes of the pure ZnSe and ZnTe crystals when the alloy composition 
changes. If we abusively apply to LO modes the classification of 
TO -mode behavior outlined in Sec.~I, we would say that the 
LO$^{+}$ mode exhibits a model type ($ii$) behavior. Certainly this is 
the origin of the MREI-VCA attribution of ZnTeSe to the type 
($ii$). Residual LO$^{\rm int.}$ and LO$^{-}$ modes, in the sense of decreasing
frequencies, are driven back towards the TO modes.

Now we turn to the TO modes (thick lines in Figs 3 and 4). We re-assign 
the presumed unique TO mode in the Raman/IR spectra of ZnTeSe as the sum of 
close $TO_{\rm Zn-Te}$ and TO$_{\rm Zn-Se}^{Se}$ modes. In fact the bi-modal 
character was already obvious in the Raman spectra recorded by Artamonov 
\textit{et al.}\cite{20} The intensities of the two modes in Fig.~4 
exhibit antagonist variations versus $x$, and become equal at
$x{\sim}$0.7, consistently with experimental observations.\cite{20} 
An additional
minor LO--TO inverted oscillator within the main TO--LO band, evidenced by 
IR reflectivity,\cite{13} is assigned as 
the LO$^{\rm int.}$--TO$_{\rm Zn-Se}^{\rm Te}$ band. The LO--TO inverted 
splitting is large at intermediate $x$ values, and further 
increases with (1$-$$x$), as evidenced by Burlakov \textit{et al.}\cite{13} 
Then it vanishes in the Te- and Se-dilute limits, as observed 
by Yang \textit{et al.}\cite{19} In brief, all the `anomalous' features in 
the Raman/IR spectra of ZnTeSe find a natural assignment within 
the percolation model, on a quantitative basis. Actually, in 
retrospect there is a quasi-perfect analogy between the phonon 
mode behaviors of ZnTeSe (this work) and InGaP (Ref.~\onlinecite{03}). 

To secure a basis of our discussion on TO modes we provide independent 
\textit{ab initio} insight into the TO density of states (TO-DOS) 
at the zone-center (ZC, ${\bf q}$=0). In principle this compares directly
to the Raman signal, neglecting however polarisability-related variations 
of intensity of different modes. We used the following expression for 
the ${\bf q}$-projected TO-DOS,
\begin{equation}
I_{\aleph}(\omega,{\bf q}) = \sum_i \left|
\sum_{\alpha{\in}{\aleph}} A_i^{\alpha}(\omega)
\exp({\bf q} {\bf R}_{\alpha})\right|^2\,,
\end{equation}
where ${\bf q}$ refers to the wavevector, $i$ stands for the cartesian 
coordinates, $A_i^{\alpha}(\omega)$ is the $i$-component of the phonon 
eigenvector for atom $\alpha$ at frequency $\omega$, 
${\bf R}_{\alpha}$ is the positional vector, and $\aleph$ represents 
an arbitrarily chosen group of atoms (say, those of a given chemical 
species). We applied Eq. (3) to a fully-relaxed (lattice parameter 
and internal coordinates) 32-atom supercell at the representative 
alloy composition $x{\sim}$0.5 (see a drawing in the inset of 
Fig.~4). Care was taken that the Se and Te atoms distribute in 
equal proportion over all substituting planes, as is expected 
in a random alloy. We apply the plane-wave pseudopotential method 
within the density functional theory and a linear response technique, 
using specifically the PWSCF method.\cite{28} The pseudopotentials 
included Zn$3d$, Se$3d$, Te$4d$ as the deepest valence states; plane 
wave cutoff of 25 Ry, Brillouin zone summation done over 2$\times$2$\times$2 
${\bf k}$-points mesh of Monkhorst and Pack,\cite{29} and local density 
approximation (LDA) used for the exchange-correlation.

The ZC TO-DOS obtained per atom for uniformly-damped Zn, Se and Te oscillators
($\sim$10~cm$^{-1}$) are displayed in the inset of Fig.~4. As expected, 
three equidistant oscillators spaced by $\sim$10~cm$^{-1}$, 
with similar intensities, show up, i.e. a ZnTe-like below two 
ZnSe-like ones (refer to the arrows).
An overall blue-shift of the ZC TO-DOS with respect to the Raman signal 
($\sim$20~cm$^{-1}$) is due to a well-known overbinding in LDA. 
We have checked that the interplay between the frequencies and intensities 
of the three oscillators in the ZC TO-DOS are consistent with 
the general trend in the TO symmetry as displayed in the body of Fig.~4,
over the entire composition domain. 

\section{`PERCOLATION' vs. `ELLIOTT-CPA' and `MREI-VCA'}
The percolation schemes for InGaAs, InGaP (see Ref.~\onlinecite{03})
and ZnTeSe are schematically reproduced in Fig.~5. These were built up 
from the experimental ($\omega_{\rm imp.}$, $\Delta$) 
values reported in table 1. Only the 
TO branches are shown, for clarity, while simplifying the rescaled-MREI 
oblique segments to straight lines. There is an obvious analogy 
between the three schemes, indicating that the traditional MREI-VCA/Elliott-CPA 
classification has, in fact, no \textit{raison d'\^{e}tre}. 

\begin{figure*}[t]
\centerline{\includegraphics[width=0.95\textwidth]{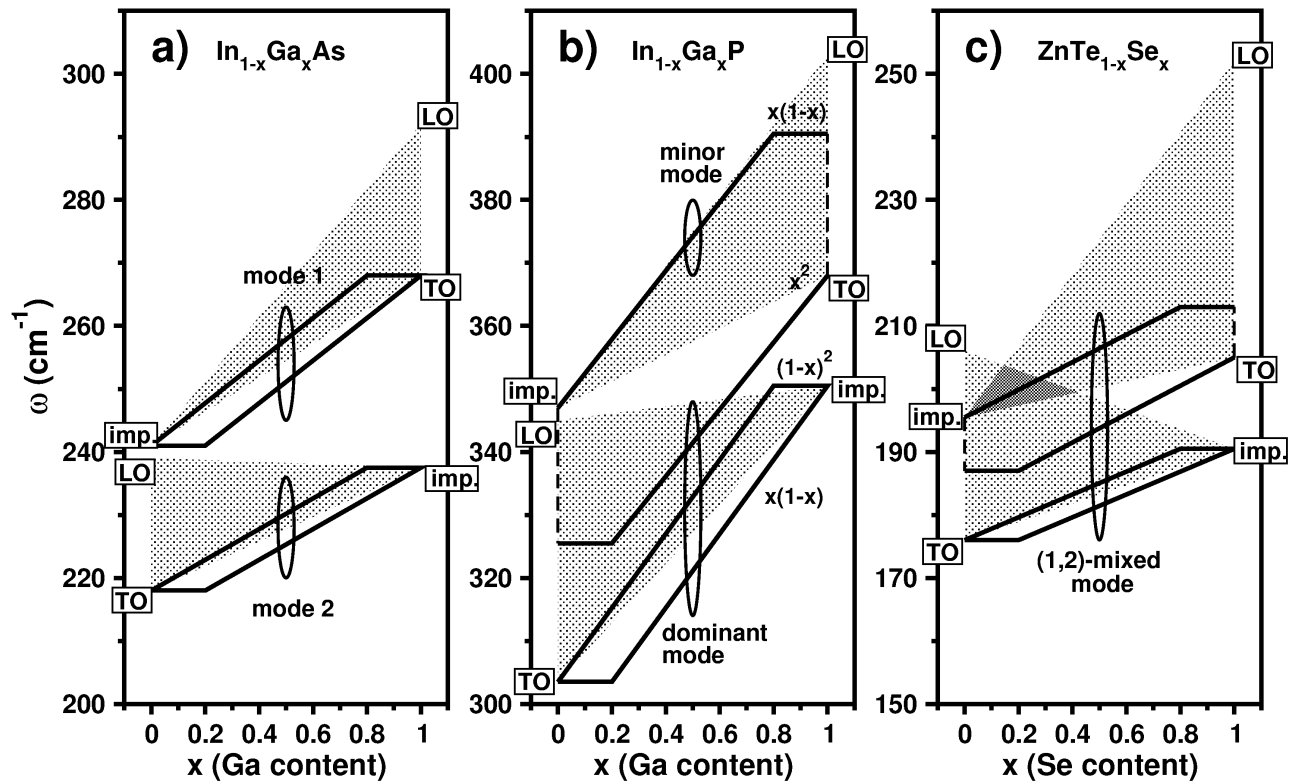}}
\medskip
\caption{%FIG. 5: 
Simplified 1-bond$\rightarrow$2-mode TO (thick lines) 
percolation schemes of InGaAs (a), InGaP (b) and ZnTeSe (c),
as built up from the experimental ($\omega_{\rm imp}$, $\Delta$) values
reported in Table~1. 
A direct comparison can be made with the MREI-VCA schemes of Fig.~2.
Dotted lines indicate the onset
of the 1-bond$\rightarrow$2-mode TO behavior just departing from
the dilute limits. The generic bond 
fraction corresponding to each TO branch (refer to Fig.~1b) is 
indicated in part b) on the right side. 
The intensity of each TO mode scales accordingly. 
The optical bands used for the Elliott-CPA criterion are shown as shaded areas. 
}
\end{figure*}

\begin{table*}[t]
\caption{Comparison of the theoretical ($\omega_{\rm imp}$, $\Delta$) values 
of the leading AB$_{1-x}$C$_x$ alloys in the MREI-VCA classification,
as derived from the \textit{ab initio} `protocol', with the experimental 
ones. A more explicit notation for $\omega_{\rm imp}$ is AB:C (AC:B), 
that refers to an isolated B (C) atom in pure AC (AB). Direct 
access is gained via Eq. (1) from the difference in the A-C (A-B) 
bond length \ensuremath{\Delta}l$_{C}$ (\ensuremath{\Delta}L$_{B}$) 
when the host medium changes from pure AC (AB) to pure AB (AC). For GaP, InP, 
ZnSe, ZnTe and GaAs the $\gamma_{\rm T}$ values are taken
from Ref.~\onlinecite{23}, for InAs from Ref.~\onlinecite{30}.
$\Delta_{\rm C}$ ($\Delta_{\rm B}$) is the splitting 
between the two like A--C (A--B) TO modes in the C-dilute (B-dilute) limit.}
\medskip
%------------------------------------------------
\begin{ruledtabular}
\begin{tabular}{ccccccccccc}
 AB$_{1-x}$C$_x$:& \multicolumn{3}{c}{In$_{1-x}$Ga$_x$As} &&
                   \multicolumn{2}{c}{In$_{1-x}$Ga$_x$P}  &&
		   \multicolumn{3}{c}{ZnTe$_{1-x}$Se$_x$} \\
\cline{2-4}\cline{6-7}\cline{9-11}		   
     & Raman/IR & EXAFS & \emph{Ab initio} &&
       Raman/IR &         \emph{Ab initio} &&
       Raman/IR & EXAFS & \emph{Ab initio} \\
\hline
 $\Delta\,l_{\rm B}$ ({\AA}) & & 2.672--2.588\footnotemark[1] & 
                                 2.637--2.584 && 
                               & 2.540--2.485 &&
                               & 2.643--2.600\footnotemark[3] & 
			         2.630--2.545 \\
 $\Delta\,l_{\rm C}$ ({\AA}) & & 2.450--2.488\footnotemark[2] & 
                                 2.460--2.495 && 
                               & 2.360--2.393 &&
			       & 2.452--2.480\footnotemark[3] & 
			         2.450--2.467 \\
 AC:B (cm$^{-1}$) & ${\sim}$237\footnotemark[1] & ${\sim}$241 & ${\sim}$234 &&
                    ${\sim}$350\footnotemark[1] & ${\sim}$330 &&
	            ${\sim}$189 & ${\sim}$191       & ${\sim}$203 \\
 AB:C (cm$^{-1}$) & ${\sim}$241\footnotemark[1] & ${\sim}$250 & ${\sim}$251 &&
                    ${\sim}$347\footnotemark[1] & ${\sim}$350 &&
                    ${\sim}$195 & ${\sim}$196       & ${\sim}$200 \\
 $\Delta_{\rm C}$ (cm$^{-1}$) & ${\sim}$0\footnotemark[1] && ${\sim}$3 &&
                    ${\sim}$22\footnotemark[1]  & ${\sim}$28 &&
		    ${\sim}$8 &       & ${\sim}$6 \\
 $\Delta_{\rm B}$ (cm$^{-1}$) & ${\sim}$0\footnotemark[1] && ${\sim}$0 &&
                    ${\sim}$0\footnotemark[1]   & ${\sim}$0  &&
		    ${\sim}$0 &       & ${\sim}$0 \\
\end{tabular}
\end{ruledtabular}
\footnotetext[1]{Ref.~\onlinecite{03}, and Refs therein;}
\footnotetext[2]{Ref.~\onlinecite{05};}
\footnotetext[3]{Ref.~\onlinecite{22}.} 
\end{table*}

Still, the percolation schemes must be consistent with the latter 
classification, which we discuss now. The Ga- and In-related TO--LO bands 
as derived according to the Elliott's procedure (see Sec.~I), do not overlap 
in InGa(As,P) (see the shaded areas in Figs 5a and 5b), while 
the Zn--Se and Zn--Te ones do overlap in ZnTeSe (see the over-shaded area 
in Fig.~5c). This is consistent with the Elliott-CPA classification 
of InGa(As,P) and ZnTeSe as types ($i$) and ($ii$), respectively. 
The percolation schemes are also consistent with the MREI-VCA classification. 
It is just a matter of regrouping close individual/double TO branches 
as shown by ovals in Fig.~5. The usual terminology of separate modes 
[type ($i$), Fig.~5a], dominant-plus-minor modes [type ($iii$), Fig.~5b] 
and mixed-mode [type ($ii$), Fig.~5c] comes out naturally. 
Note that in the percolation schemes the individual TO branches
all remain quasi-parallel, as ideally expected.

A probable reason why differences in the detail of the phonon 
behaviors of InGaAs, InGaP and ZnTeSe were previously mistaken 
for differences in the principles of the phonon mode behaviors 
lies in an implicit, but wrong, assumption that the TO and LO modes behave 
similarly in an alloy. In the LO symmetry a strong $\vec{E}$-coupling occurs
in particular
between the like LO modes that come from the same double-branch, because 
these refer to the same bond species and as such have close frequencies 
in general. The result is that the actual 1-bond$\rightarrow$2-mode behavior 
visible in the TO symmetry, supporting a description of the phonon mode 
behavior at the \textit{mesoscopic} scale on a percolation basis, is literally 
rubbed out in the LO symmetry. What is left is an apparent 
\mbox{1-bond$\rightarrow$1-mode} LO behavior, encouraging 
a more crude description 
at the \textit{macroscopic} scale according to the MREI-VCA. Due to this 
TO vs. LO difference in nature, any attempt to explain the TO and LO
modes in an alloy on the same intuitive basis is damned to fail. 
Simplicity arises by focusing on TO modes; LO modes follow.

\section{\textit{AB INITIO} `PROTOCOL' FOR A SELF-SUFFICIENT 
PERCOLATION MODEL}
To complete the picture, we give in this section a simple \textit{ab initio} 
`protocol' to estimate the 
basic input parameters of our semi-empirical percolation model, 
i.e. ($\omega_{\rm imp.}$, $\Delta$) per bond. The `protocol' operates 
at the impurity-dilute limits, for two reasons. First, it is 
attractive conceptually to be in a position to derive the whole 
Raman/IR behavior of an alloy from its behavior in the dilute 
limits. Incidentally, the Elliott's criterion proceeds from a 
similar ambition. Second, placing the analysis at the dilute 
limits brings in a decisive advantage that the number of motifs 
is much restricted for the impurity atoms. Then, it is just a matter 
to identify those motifs that are suitable for our purpose. Further, 
at the impurity limit a rather small-size supercell will do, 
in which case a full \textit{ab initio} approach can be pursued. Beyond 
the impurity limit, we would face a necessity to incorporate 
the whole statistics of the alloy disorder, averaging over a 
big number of large-size supercells. Almost invariably that would 
mean employing a simplified (e.g. an empirical valence force 
field) method to perform structure relaxations. Such calculations 
have been done, e.g. in Refs \onlinecite{21} and \onlinecite{31}.

Our `protocol' is based on \textit{ab initio} bond length/phonon calculations 
in prototype 64-atom (2$\times$2$\times$2-replicated simple cubic) supercells 
of zincblende structure, with either one or two nearest cation 
(or anion) sites substituted by a different species. Calculation 
of zone-center phonons was preceded by unconstrained 
(lattice parameter and atom positions) relaxation for each supercell. 
In particular the sufficiency of the supercell size for our purpose 
of treating impurity pairs can be supported by the following 
arguments. First, some of us checked the falling down of interatomic 
force constants in semiconductor alloys with distance, and found 
that the interactions beyond the second neighbors can be safely 
discarded, without noticeable differences in the phonon spectra.\cite{32} 
In our supercell with an impurity pair, none of the second neighbors 
of an impurity is simultaneously a second neighbor to a spurious 
(translated) impurity. Second, even with the break of symmetry 
introduced by an impurity pair, the relaxed shape of the supercell 
remains remarkably cubic. Finally Teweldeberhan \textit{et al}.\cite{33} 
who were confronted with a similar problem of \textit{ab initio} (bond 
length) calculations related to chosen (In,N) motifs in the highly-dilute 
(In,N)-impurity limit of In$_y$Ga$_{1-y}$As$_{1-x}$N$_x$ have observed 
that the 64-atom supercell calculation is well-converged in size 
for such purpose, i.e. the 64-atom supercell seems sufficient 
to reproduce the behavior of the considered motif as immersed 
in the infinite solid. The LDA calculations have been done in 
part (ZnTeSe) with the PWSCF method,\cite{28} as stated above, and 
in part (InGaAs, InGaP) with {\sc Siesta},\cite{34} which relies 
on norm-conserving pseudopotentials and strictly confined atom-centered 
numerical basis functions. The basis sets (from As$3d$, Ga$3d$ upwards) 
were of `double-zeta with polarization orbitals' quality. Brillouin 
zone integration demanded at least 2$\times$2$\times$2-mesh 
in the ${\bf k}$-space, in order to get convergence of phonon frequencies 
with respect to this parameter.

To access $\omega_{\rm imp.}$ we use a supercell containing a single 
impurity ($\sim$3\% Imp.). This is the ultimate configuration 
that refers to an impurity in the region rich of the other substituting 
species, i.e. the dominant one. We search for the impurity bond 
length and estimate the difference ${\Delta}l$ with respect to 
the bond length in the pure crystal. Eventually $\omega_{imp.}$ is derived 
via $\gamma_{\rm T}$ by using Eq.~(1). The directly calculated impurity 
frequency $\omega_{\rm imp.}^{\rm calc.}$ is not necessary at this stage 
as it might be subject to systematic shift, due to the effective overbinding 
caused by the LDA. The as-obtained $\omega_{\rm imp.}$ values for InGaAs, 
InGaP and ZnTeSe fairly agree with the experimental ones, as shown 
in table~1, which validates this part of the `protocol'.

To access $\Delta$, we use a supercell containing two neighboring 
substitutional impurities. This pair forms the germ of the impurity-rich 
region, i.e. the ultimate configuration that refers to an impurity 
staying in its own environment. We search for the frequency of 
the softer impurity-related phonon mode (see detail further on), 
referred to as $\omega_{\rm pair}^{\rm calc.}$, and $\Delta$ is estimated as 
$\left|\omega_{\rm imp.}^{\rm calc.}-\omega_{\rm pair}^{\rm calc.}\right|$.
The systematic error due to LDA is thus eliminated.

\begin{figure}[tb]
\centerline{\includegraphics[width=0.48\textwidth]{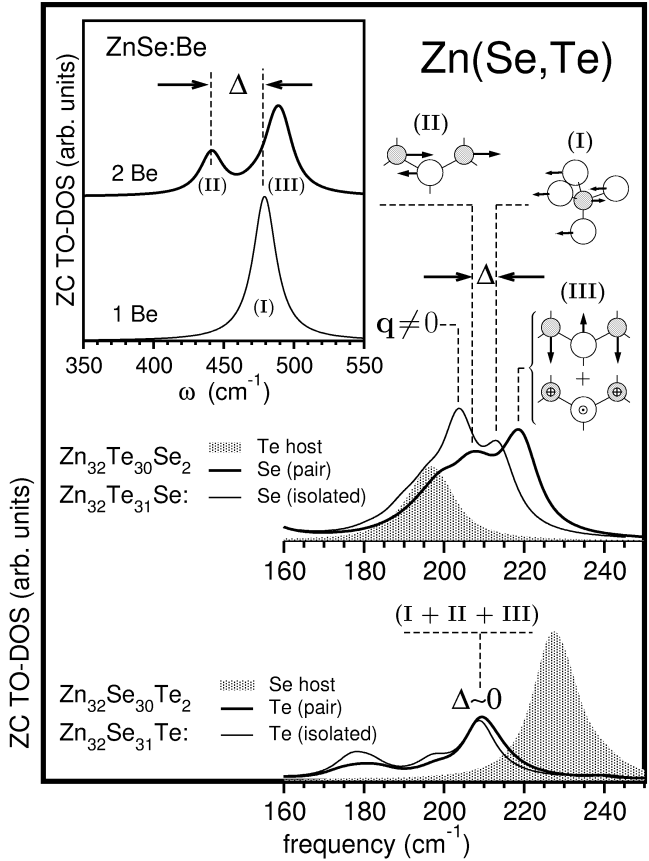}}
\medskip
\caption{%FIG. 6: 
ZC TO-DOS per atom related to the dominant substituting 
species (shaded areas), to the isolated impurity (thin lines)
and to the impurity pair (plain lines) 
at the Se- (upper curves) and Te-dilute (lower curves) limits of simple cubic 
64-atom ZnTeSe supercells. The vibration patterns of the impurity
modes are indicated with labeling (I) to 
(III), the impurity (Zn) species 
being represented by filled (open) symbols. The splitting between 
the mode due to the isolated impurity and the softer mode 
of the impurity-pair provides a direct estimate of $\Delta$ 
(one of the two input parameters of the percolation model), as indicated
by the antagonist arrows. The 
notation `${\bf q}{\neq}0$' indicates a parasitical vibration mode 
out of the center of the Brillouin zone. In the inset, the 
ZC TO-DOS per atom due to the isolated Be impurity (upper curve) and to 
the Be impurity pair (lower curve) in ZnSe are shown, for reference
purpose.}
\end{figure}

A choice system to test the `protocol' with respect to $\Delta$ 
is the random Zn$_{1-x}$Be$_{x}$Se alloy, that exhibits an unequally 
well-resolved 1-bond$\rightarrow$2-mode Raman behavior in the Be--Se spectral 
range.\cite{03} We report in the inset of Fig.~6 the ZC TO-DOS curves 
per-atom related to the single Be impurity (lower curve) 
and to the Be impurity pair (upper curve). By forming the pair, 
the original triply degenerate impurity mode splits into two, 
a doubly-degenerate mode at a slightly higher frequency corresponding 
to anti-phase Be and Se vibrations transverse to the Be-Se chain 
(in-plane and out-of-plane ones), and a minor one, red-shifted 
by $\sim$40~cm$^{-1}$ (the softer pair-related phonon mode mentioned 
above), due to similar vibrations but along the chain, i.e. longitudinal 
ones then. Somewhat ideally, a similar shift $\Delta$ of $\sim$40~cm$^{-1}$ 
is observed between the two like Be-Se Raman modes 
at small Be content.\cite{35} 
Only, the theoretical features are globally blue-shifted 
by $\sim$35~cm$^{-1}$, due to the effective overbinding in LDA. 

Now we apply the same procedure to ZnTeSe. The ZC TO-DOS due to the isolated 
impurity (thin lines) and to the impurity pair (thick lines) at both the Se- 
(upper curves) and Te-dilute (lower curves) limits are shown in the body 
of Fig.~6. The ZC TO-DOS due to the dominant substituting species 
(shaded areas) are added for complementary insight into the reference 
TO mode of the host lattice. Again, the theoretical curves 
are globally blue-shifted (by $\sim$20~cm$^{-1}$) 
with respect to the Raman/IR features, 
due to the LDA. At the Se-dilute limit, two well-resolved pair-impurity 
(Se) modes show up with the isolated-impurity (Se) mode in-between, 
as in the reference ZnBeSe system. The host-lattice (Te) mode 
is situated just below. The low-frequency pair-impurity mode 
is red-shifted by $\Delta{\sim}$6~cm$^{-1}$ with respect to 
the isolated-impurity mode. Note that the dominant contribution 
in the ZC TO-DOS of the isolated Se impurity does not refer to a pure ZC mode. 
Indeed we have checked that the impurity Se atom does not vibrate 
against the whole cage of its Zn first-neighbors, two out of 
four Zn atoms remain quasi-immobile. At the Te-dilute limit, 
the pair-impurity (Te) modes, transverse and longitudinal to 
the Te-Zn-Te chain, do nearly degenerate into a single mode situated 
at the same frequency as the isolated-impurity (Te) mode, corresponding 
to $\Delta{\sim}$0~cm$^{-1}$. The host-lattice (Se) mode is 
located at a slightly higher frequency. These are precisely the 
limit (isolated impurity, impurity-pair, host lattice) configurations 
that we expect at $x{\sim}$(0,1) from Fig.~3. Similar insight 
into the $\Delta$ values of In$_{1-x}$Ga$_x$As and In$_{1-x}$Ga$_x$P 
is detailed in Ref.~\onlinecite{35}. 

The $\Delta$ values obtained via the `protocol' for 
InGaAs, InGaP and ZnTeSe are compared to the experimental ones 
in table~1. Again, the agreement is rather good, which validates 
the second part of the `protocol'. 

For sake of completeness we mention also a similar procedure in the N-dilute 
limit of the \textit{non-random} GaAs$_{1-x}$N$_x$ alloy.\cite{36} 
Again the contrast in the bond length/stiffness of the two constituent 
species is large, even larger than in ZnBeSe, which provides 
a large $\Delta$ value for the Ga-N bond. The theoretical estimate 
$\Delta{\sim}$50~cm$^{-1}$ compares reasonably well to the experimental value 
of $\sim$40~cm$^{-1}$, as inferred from the Raman spectra. Note that in this 
highly-contrasted system the fine structure of the ZC TO-DOS in the spectral 
region of the isolated N impurity and of the harder mode of the N-impurity pair
is completely resolved, i.e. the in-plane and out-of-plane transverse modes 
of the impurity pair are no more degenerate.

Together with the \textit{ab initio} `protocol' our semi-empirical 
percolation model gives a self-sufficient tool for an insight 
into the vibration spectra of, in principle, any zincblende alloy. 

\section{CONCLUSION }
In summary, by putting an emphasis on TO modes -- as opposed to 
LO modes -- as the proper way to get reliable insight into the whole 
complexity of the phonon mode behavior of an alloy, we show that 
ZnTeSe obeys the 1-bond$\rightarrow$2-phonon percolation model, as InGaAs 
and InGaP do. We propose a three-oscillator 
[1$\times$(Zn--Te), 2$\times$(Zn--Se)] version, independently supported 
by existing EXAFS data and home-made \textit{ab initio} phonon/bond length 
calculations. This leads to unification of the traditional 
MREI-VCA/Elliott-CPA classification into a single class covered by 
the percolation model. Also, this work reveals that TO-based vibrational 
spectroscopies provide natural insight (non destructive and contactless) 
into the alloy disorder at the unusual mesoscopic scale, 
which is hardly achieved 
otherwise. In particular, this offers an attractive perspective for the study 
of the long scale (self) organization in alloys, be it local ordering\cite{03}
or anti-clustering.\cite{36} At last we propose a simple \textit{ab initio} 
`protocol' at the dilute-impurity limits to estimate the input parameters 
of the semi-empirical percolation model for the calculation 
of the Raman/IR spectra of a random zincblende alloy. With this, 
the model becomes self-sufficient. More generally, what emerges is that 
SC alloys can not escape a description of some of their very basic physical
properties via a percolation concept. As such, they do not differ 
fundamentally from the considerably more complex molecular/natural mixtures. 

\begin{acknowledgments}
The authors acknowledge support from the Indo-French Center for 
the Promotion of Advanced Research (IFCPAR project N\ensuremath{^\circ} 
3204-1) and from the \textit{Centre National Informatique de l'Enseignement 
Sup\'{e}rieur} (CINES project N\ensuremath{^\circ} pli2623).
\end{acknowledgments}

\end{document}